\documentclass[twocolumn,showpacs,preprintnumbers,amsmath,amssymb,superscriptaddress]{revtex4}
\usepackage{graphicx,bm}
\def\be{\begin{equation}}
\def\ee{\end{equation}}
\def\bea{\begin{eqnarray}}
\def\eea{\end{eqnarray}}

\def\bbuildrel#1_#2^#3{\mathrel{\mathop{\kern 0pt#1}\limits_{#2}^{#3}}}
\newcommand{\gsim}{\;\rlap{\lower 3 pt \hbox{$\mathchar \sim$}} \raise 2pt \hbox {$>$}\;}
\newcommand{\lsim}{\;\rlap{\lower 3 pt \hbox{$\mathchar \sim$}} \raise 2pt \hbox {$<$}\;}

\def\theequation{\arabic{section}.\arabic{equation}}
\newcommand{\newsection}[1]{\section{#1}\setcounter{equation}{0}}
\newcommand{\newappendix}[1]{\section*{#1}\setcounter{equation}{0}}

\begin{document}
\title{Phase space analysis for  three and four massive particles in  final states}
\author{H.M.~Asatrian}
\author{A.~Hovhannisyan}
\author{A.~Yeghiazaryan}
\affiliation{Yerevan Physics Institute, 0036 Yerevan, Armenia}
\begin{abstract}
We propose  formulae for computing the phase space integrals of
$1\to 3$ and $1\to 4$ processes with  massive particles in final
states. As an application of these formulae we study the final state
mass effects in some interesting phenomenological cases, giving
fully integrated analytic results for the corresponding phase
spaces. We consider also the $B_s-\bar{B}_s$ process at NNLO and
calculate one of the most complicated master integrals, which
contributes to the $\Delta\Gamma_{B_s}$ at $O(\alpha_s^2)$.
\end{abstract}
\maketitle

\newsection{Introduction \label{sec:intro}}
The $B_{s,d}$  -meson rare decays and oscillations are known to be a
unique source of indirect information about physics at scales of
several hundreds GeV. In the Standard Model all these processes
proceed through loop diagrams and thus are relatively suppressed. In
the extensions of the Standard Model the contributions stemming from
the diagrams with "new" particles in the loops can be comparable or
even larger than the contribution from the SM. Thus getting
experimental information about rare decays puts strong constraints
on the extensions of the SM or, if we are lucky, can lead to a
strong disagreement with the SM predictions providing an evidence of
some "new physics". To make a rigorous comparison between experiment
and theory, one has to get refined theoretical predictions for the
rare decay at hand. In particular,  perturbative QCD corrections
through next-to-next-to leading order (NNLO) in $\alpha_S$ are
needed. When calculating higher order QCD corrections, along with
the calculation of virtual corrections, it is necessary to take into
account real emission of gluons and quark-antiquark pairs. Then the
phase space integrals contain infrared and collinear singularities,
which can be regularized using dimensional regularization scheme.

Sometimes however it is more convenient to introduce a small mass as
an infrared regulator which can regularize both the soft and
collinear divergences. This can be for instance massive s quark for
$b\to s \gamma$ \cite{Kapustin:1995fk} and $b\to s \gamma \gamma$
\cite{Asatrian:2011ta}  or massive light quarks for $b\to
s\bar{q}q\gamma$ $(q=u, d, s)$ \cite{Kaminski:2012eb}. Another
example is the calculation of CP-asymmetry and width
difference-$\Delta\Gamma$ in $B_{s,d}-\bar{B}_{s,d}$ mixing at NLO
and NNLO in $\alpha_s$, when the calculation gets simplified by the
usage of a small gluon mass as an infrared regulator.

In this article we derive formulae for the $1\to 3$
\cite{Scharf:1993ds} and $1\to 4$ phase spaces, where the final
state particles are massive. Following the approach of
\cite{Asatrian:2002va} for the $1\to 3$ process  we derive formulae
for the phase space in the rest frame of two particles in the final
state and obtain a factorized parameterization of the  phase-space
for the case when all three particles in final state are massive. In
case of $1\to 4$ process we consider similar method, deriving the
phase space formula in the rest frame of three particles in the
final state. Here we come to the factorized parameterization of
phase-space  when only one particle in the final state is massive.

The paper is organized as follows. In Section II we discuss
three-particle phase space. Section III is devoted to the
investigation of  four-particle phase space. In Section IV we apply
our technique to the calculation of master integral connected with
the $B-\bar{B}$ mixing at $O(\alpha_s^2)$. The Appendix contains
formulae of $1\to 3$ and $1\to 4$ integrated phase spaces for some
particular cases.

\newsection{Three-Particle Phase Space}

Here we consider the three particle decay, when particle with
momentum $p^{'}$ decays into three particles with momenta $p^{'}_i$
and masses $m_i$, $i=1,2,3$; $p^{'}=p^{'}_1+p^{'}_2+p^{'}_3$. We
start from the well-known expression for the differential decay
width:
\begin{eqnarray}
\label{phN} \nonumber d\Gamma=\frac{1}{2m} \overline{|M|^2} D\Phi \,
,
\end{eqnarray}
where $\overline{|M|^2}$ is the squared matrix element, summed and
averaged over spins and colors of the particles in the final and
initial states respectively and $m$ is the mass of decaying
particle.

The phase space formula in the rest frame of particles 1 and 3 with
momenta $p_1^{'}$ and $p_3^{'}$ was derived in
\cite{Asatrian:2002va}:
\begin{eqnarray}
\label{phasefactorsleptrest}
&& \hspace{-0.3cm} D\Phi^{}=m^{2(d-3)} D\Phi^{}_1\, D\Phi_2 \, ds_{13}\, ,\\
&&\nonumber \hspace{-0.3cm} D\Phi^{}_1= (2\pi)^d \,
s_{13}^{(d-2)/2}\, \frac{d^{d-1} p}{2 p^0}
\frac{d^{d-1}p_2}{(2\pi)^{d-1}2p_2^0} \, \delta^d(p-p_2-q),\\
\nonumber && \hspace{-0.3cm} D\Phi_2=
\frac{d^{d-1}p_1}{(2\pi)^{d-1}2p_1^0} \,
\frac{d^{d-1}p_3}{(2\pi)^{d-1}2p_3^0} \, \delta^d(q-p_1-p_3) \, .
\end{eqnarray}
where $d=4-2\epsilon$ and   "dimensionless" momenta $p=p^{'}/m$,
$p_i=p_i^{'}/m$, i=1,2,3 are introduced. In
(\ref{phasefactorsleptrest}) we introduced an additional integration
over $s_{13}=q^{2}$, $q=p_1+p_3$ (in the considered frame
$q=(\sqrt{s_{13}},\vec{0})$ ). Now we can integrate over  $d-1$
components of $p_1$ and $p_2$ using the spatial parts of  two
$d$-dimensional $\delta$ functions. To carry out the remaining
integrations we choose the coordinate axes in a way that  particle
momenta have the following components:
\begin{eqnarray} \label{vectorform}
p &=& (E,|\vec{p}|,0,0,....) \, , \nonumber \\
p_3 &=& (E_3,|\vec{p}_3| \cos \vartheta, |\vec{p}_3| \sin
\vartheta,0,....)\, ,
\end{eqnarray}
where the dots correspond to the components of extra space
dimensions, which are all zero. Making use of the remaining two
one-dimensional $\delta$-functions we  express $E$ and $E_3$ in the
following way:
\begin{eqnarray}
E = \frac{1+s_{13}-x_2}{2\sqrt{s_{13}}} \quad ; \quad E_3 =
\frac{x_3+s_{13}-x_1}{2\sqrt{s_{13}}} \, ,
\end{eqnarray}
where $x_i=m^2_i/m^2$, $i=1,2,3$ are dimensionless variables. After
integration over those angles on which $\overline{|M|^2}$ does not
depend, we obtain a factorized formula for the three massive
particles phase space (unlike \cite{Scharf:1993ds}, where the
corresponding formula is not factorized):
\begin{widetext}
\begin{eqnarray}
\label{PS3} \nonumber D\Phi(1 \to 3) &=& m^{2(d-3)}
\frac{\Omega_{d-1} \, \Omega_{d-2} }{2^{4} \, (2\pi)^{2d-3} }
\,s_{13}^{(d-4)/2} \, \left(|\vec{p}| \, |\vec{p}_3| \,
\sin\vartheta \right)^{d-3} \, d\vartheta \, ds_{13}
\\
&=& m^{2(d-3)} \frac{\Omega_{d-1} \, \Omega_{d-2} }{2^{4} \,
(2\pi)^{2d-3} } \,s_{13}^{(d-4)/2} \, \left(|\vec{p}| \, |\vec{p}_3|
\, \right)^{d-3} (1-\cos^2\vartheta)^{\frac{d-4}{2}} \,
d\cos\vartheta \, ds_{13}
\end{eqnarray}
where $\Omega_d$ is the solid angle in $d$ dimensions,
\end{widetext}
\begin{eqnarray}
\Omega_d=\frac{2\pi^{d/2}}{\Gamma(d/2)}\,,
\end{eqnarray}
and
\begin{eqnarray}
\nonumber && |\vec{p}| =
\frac{1}{2\sqrt{s_{13}}}\sqrt{(1+x_2-s_{13})^2-4x_2}\, ,
\\
\quad && |\vec{p}_3| =
\frac{1}{2\sqrt{s_{13}}}\sqrt{(x_1+x_3-s_{13})^2-4x_1x_3} \,
\end{eqnarray}
are 3-momenta expressed by $K\ddot{a}ll\acute{e}n$ functions.

For the integration limits we have
\begin{eqnarray}
&& \nonumber (\sqrt{x_1}+\sqrt{x_3}\,)^2\leq s_{13}\leq
(1-\sqrt{x_2}\,)^2 ~,
\\
&& \hspace{1.5cm} -1\leq \cos\vartheta \leq1.
\end{eqnarray}

We represent all  scalar products via variables $s_{13}$ and
$\cos\vartheta$:
\begin{eqnarray}
&& \nonumber p.p_2=\frac{1+x_2-s_{13}}{2}
\\
&& \nonumber p.p_1=\frac{2(1+x_1+x_3)-x_2-X}{4}
\end{eqnarray}
\begin{eqnarray}
&& \nonumber p.p_3=\frac{-2(x_1+x_3)-x_2+2s_{13}+X}{4}
\\
&& \nonumber p_2.p_1=\frac{2+4x_3-x_2-2s_{13}-X}{4}
\\
&& \nonumber p_2.p_3=\frac{-x_2-4x_3+X}{4}
\\
&& p_1.p_3=\frac{s_{13}-x_1-x_3}{2}
\end{eqnarray}
where
\begin{eqnarray}
\nonumber X&=& -4|\vec{p}| \,
|\vec{p}_3|\cos\vartheta+1+2(x_1+x_3)-s_{13}
\\ &&-\frac{(1+s_{13}-x_2)(x_1-x_3)}{s_{13}}\,,
\end{eqnarray}
and $|\vec{p}|$ and $|\vec{p}_3|$ are defined in (2.6).

From the formula (\ref{PS3}) for the three-particle phase space one
can get a simpler formula assuming that  the masses of two particles
in the  final state are equal to zero: $m_{2}=m_{3}=0$. In that case
we make the replacements: $\cos\vartheta=2\lambda_1-1$ and
$s_{13}=\lambda_2 (1-x_1)+x_1$, where $0<\lambda_{1,2}<1$. Then we
get:
\begin{widetext}
\begin{eqnarray}
\label{PS3N} D\Phi(1 \to 3) &=& m^{2(d-3)} \frac{\Omega_{d-1} \,
\Omega_{d-2} }{2^{d+1} \, (2\pi)^{2d-3} }\, d \lambda_1 \,
d\lambda_2
\\
&& \nonumber \times \, (1-x_1)^{2d-5}\,\left [\lambda_2 (1-x_1)+x_1
\right ]^{(2-d)/2} \, \left[\lambda_2\, (1-\lambda_2)\right]^{d-3}\,
\left[\lambda_1(1-\lambda_1)\right]^{\frac{d-4}{2}}
\end{eqnarray}
\end{widetext}
If we put $x_1=0$ (i.e. $m_1=0$) in the expression (\ref{PS3N}), we
get the formula (2) of \cite{Anastasiou:2003gr}.

In the Appendix A we study the final state mass effects in some
interesting phenomenological cases, providing fully integrated
analytic formulae for the $1\to 3$ processes.

\newsection{Four-Particle Phase Space}

Now we proceed to the case of four particle decay. For this case one
can derive the expression for the phase space  in the similar way as
in \cite{Asatrian:2002va}. The only difference is that instead of
two particle's rest frame introduced in \cite{Asatrian:2002va} here
the rest frame of three particles is used (see \cite{GehrmannDe
Ridder:2003bm}). We consider a particle with momentum $p^{'}$, which
decays into four particles with momenta $p_i^{'}$ and masses $m_i$,
$i=1,...,4$ and $p^{'}=p_1^{'}+p_2^{'}+p_3^{'}+p_4^{'}$. Here also
we introduce "dimensionless" momenta $p=p^{'}/m$, $p_i=p_i^{'}/m$,
i=1,...,4. In the rest frame of particles 2, 3 and 4 with momenta
$p_2^{'}$, $p_3^{'}$, $p_4^{'}$ (where
$\vec{p}_2+\vec{p}_3+\vec{p}_4=0$) four-particle phase space can be
written in the following way
\begin{eqnarray}
&&d\Phi^{}=m^{3d-8}(2\pi)^dd\Phi^{}_1\, d\Phi_2 \, ds_{234}\, ,\\
&&\nonumber d\Phi^{}_1=(s_{234})^{(d-2)/2} \, \frac{d^{d-1} p}{2
p^0} \, \frac{d^{d-1}p_1}{(2\pi)^{d-1}2p_1^0} \, \delta^d(p-p_1-q)\,
,
\\
\nonumber &&D\Phi_2=\frac{d^{d-1}p_2}{(2\pi)^{d-1}2p_2^0} \,
\frac{d^{d-1}p_3}{(2\pi)^{d-1}2p_3^0} \,
\frac{d^{d-1}p_4}{(2\pi)^{d-1}2p_4^0} \,
\\ \nonumber && \hspace{1cm} \times \delta^d(q-p_2-p_3-p_4)
\, ,
\end{eqnarray}
where $s_{234}$=$(p_2+p_3+p_4)^2$ and $q$ is a d-dimensional vector
with components $(\sqrt{s_{234}},\vec{0})$ in the rest frame of
particles 2, 3 and 4. We choose the coordinate axes in such a way,
that the particle momenta have the following components:
\begin{eqnarray} \label{vectorform}
&& \hspace{-0.45cm} p_2 = (E_2,|\vec{p}_2|,0,0,....) \, , \nonumber
\\
&& \hspace{-0.45cm} p_1 = (E_1,|\vec{p}_1| \cos \vartheta_1,
|\vec{p}_1| \sin \vartheta_1,0,....)\, ,
\\
&& \hspace{-0.45cm} p_3 = (E_3,|\vec{p}_3| \cos \vartheta_3,
|\vec{p}_3| \sin \vartheta_3\cos \phi_3,|\vec{p}_3| \sin
\vartheta_3\sin \phi_3,....) , \nonumber
\end{eqnarray}
where the dots, as before, correspond to the components of extra
space dimensions, which are all zero. We can immediately integrate
over those angles on which none of the scalar products depends and
obtain
\begin{eqnarray}
&& d^{d-1}p_2= \Omega_{d-1}\,|\vec{p}_2|^{d-2}~d|\vec{p}_2|\,,
\\ \nonumber
&& d^{d-1} p_1=
\Omega_{d-2}\,(\sin\vartheta_1)^{d-4}~d\cos\vartheta_1~|\vec{p}_1|^{d-2}~d|\vec{p}_1|\,,~~~~~
\\
&& \nonumber d^{d-1}p_3=
\Omega_{d-3}\,(\sin\phi_3)^{d-5}~d\cos\phi_3~(\sin\vartheta_3)^{d-4}
\\ \nonumber && \hspace{1.3cm} \times d\cos\vartheta_3~|\vec{p}_3|^{d-2}~d|\vec{p}_3|\,.
\end{eqnarray}
$p_4$ and $p$ can be integrated out immediately using the spatial
parts of the $\delta$ functions in (3.1). Thus for the phase space
we get
\begin{widetext}
\begin{eqnarray}
&& \hspace{-0.6cm} D\Phi_1=  \frac{(s_{234})^{(d-2)/2}}{2E} \,
\frac{\Omega_{d-2}(\sin\vartheta_1)^{d-4}~d\cos\vartheta_1~|\vec{p}_1|^{d-2}~d|\vec{p}_1|}{(2\pi)^{d-1}2E_1}
\, \delta\left(E-E_1-\sqrt{s_{234}}\right),
\\
&& \hspace{-0.6cm} D\Phi^{}_2=
\frac{\Omega_{d-1}|\vec{p}_2|^{d-2}d|\vec{p}_2|}{(2\pi)^{d-1}2 E_2}
\, \frac{\Omega_{d-3}
(\sin\phi_3)^{d-5}d\cos\phi_3(\sin\vartheta_3)^{d-4}d\cos\vartheta_3
|\vec{p}_3|^{d-2}d|\vec{p}_3|}{(2\pi)^{d-1}2E_3}
\frac{\delta\left(\sqrt{s_{234}}-E_2-E_3-E_4\right)\,}{(2\pi)^{d-1}2E_4},
\,
\end{eqnarray}
where
\end{widetext}
\begin{eqnarray}
&&E_4=
\sqrt{|\vec{p}_2|^2+|\vec{p}_3|^2+2|\vec{p}_2||\vec{p}_3|\cos\vartheta_3+x_4}\,,~~~~~
\\
&&E=\sqrt{|\vec{p}_1|^2+1}\, ,
\\&& x_i=m_i^2/m^2, i=1,...,4   \nonumber
\end{eqnarray}
where $|\vec{p_i}|=\sqrt{E_i^2-x_i}$, $i=1,2,3$. Making use of the
remaining two one-dimensional $\delta$-functions, we can express
$E_1$ and $E_3$ in terms of the other variables. Using the $\delta$
function in $d\Phi_1$ and (3.7) we get:
\begin{eqnarray}\label{E1E3}
E_1 = \frac{1-x_1-s_{234}}{2\sqrt{s_{234}}} \,\, , \quad E =
\frac{1-x_1+s_{234}}{2\sqrt{s_{234}}} \, .
\end{eqnarray}
From the $\delta$ function in $d\Phi_2$ together with (3.6) we find
the expression for $E_3$
\begin{widetext}
\begin{eqnarray}\label{theta1}
\nonumber\ E_3&=&\frac{1}
   {-2\left(\sqrt{s_{234}}-E_2\right)^2+2 |\vec{p_2}|^2 \cos ^2 \vartheta_3}\,\left[\left(E_2-\sqrt{s_{234}}\right)
   \left(s_{234}-2 E_2 \sqrt{s_{234}}+x_2+x_3-x_4\right)\right.
   \\
  &+& \left.|\vec{p}_2| \cos \vartheta_3 \sqrt{\left(s_{234}-2 E_2
\sqrt{s_{234}}+x_2-x_3-x_4\right)^2-4 x_3
\left(x_4+|\vec{p}_2|^2\sin ^2 \vartheta_3)\right)}\right].
\end{eqnarray}
Finally, introducing new variables: $z_1=\cos\vartheta_1$,
$z_{31}=\cos\vartheta_3$, $z_{32}=\cos\phi_3$, we get for the
four-particle phase space:
\begin{eqnarray}\label{PS4}
\nonumber D\Phi(1\to 4)&=&\frac{m^{3d-8}}{32} \,
\frac{\Omega_{d-1}\Omega_{d-2}\Omega_{d-3}}{(2\pi)^{3d-4}} \,dz_1
~dz_{31}~dz_{32}~dE_2~ds_{234}
\\
&\times&
(1-z_1^2)^{\frac{(d-4)}{2}}~(1-z_{31}^2)^{\frac{(d-4)}{2}}~(1-z_{32}^2)^{\frac{(d-5)}{2}}
{\frac{
W^{d-3}}{\sqrt{s_{234}}-E_2+{\frac{E_3}{|\vec{p_3}|}|\vec{p_2}|~z_{31}}}}
\end{eqnarray}
\end{widetext}
where
\begin{eqnarray}
W=|\vec{p_1}|~|\vec{p_2}|~ |\vec{p_3}|~ \sqrt{s_{234}}\,.
\end{eqnarray}
The integration variables  $s_{234},E_2,z_1,z_{31},z_{32}$ in
(\ref{PS4})
 have the following limits
\begin{eqnarray}
(\sqrt{x_2}+\sqrt{x_3}+\sqrt{x_4})^2\leq s_{234}\leq
(1-\sqrt{x_1}\,)^2\,,
\end{eqnarray}
\begin{eqnarray}
\nonumber &&\sqrt{x_2}\leq E_2\leq \frac{s_{234}+x_2-(\sqrt{x_3}+
\sqrt{x_4})^2}{2\sqrt{s_{234}}}\,,
\\
&& \hspace{2cm} -1\leq z_i\leq1\,.
\end{eqnarray}

Different possible scalar products between 4-momenta of the
particles can be obtained automatically from the expressions for
their components (3.2) using  the conservation of momentum. All the
scalar products can be expressed through the 5 integration variables
in (\ref{PS4}):
\begin{eqnarray} && \nonumber
p_1.p_2=E_1E_2-|\vec{p_1}||\vec{p_2}|\cos{\vartheta_1}\,,
\\
&& p_2.p_3=E_2E_3-|\vec{p_2}||\vec{p_3}|\cos{\vartheta_3}\,,
\\
&& \nonumber p_1.p_3 =
E_1E_3-|\vec{p_1}||\vec{p_3}|(\cos{\vartheta_1}\cos{\vartheta_3}
\\ \nonumber && \hspace{1cm} +\sin{\vartheta_1}\sin{\vartheta_3}\cos{\phi_3})\,,
\end{eqnarray}
and
\begin{eqnarray}
&&\nonumber \hspace{-1cm} p_1.p_4 =
E_1\sqrt{s_{234}}-p_1.p_3-p_1.p_2\,,
\\
&&\nonumber \hspace{-1cm} p_2.p_4=E_2\sqrt{s_{234}}-x_2-p_2.p_3 \,,
\\
&&\hspace{-1cm} p_3.p_4 = E_3\sqrt{s_{234}}-x_3-p_2.p_3\,,
\end{eqnarray}
In right-hand side of (3.15) the expressions given in (3.14) should
be used. Note that in (3.14) and (3.15) $E_1$ and $E_3$ are
determined by formulae (3.8) and (3.9).

We can also obtain a factorized formula for the 4 particle phase
space with 1 massive and 3 massless particles in the final state,
taking $x_{2}=x_{3}=x_{4}=0$. Making the following substitutions in
(\ref{PS4})
\begin{eqnarray}
\nonumber && s_{234}=(1-\sqrt{x_1})^2 \lambda_1,\,
E_2=\frac{1}{2}\sqrt{\lambda_1}(1-\lambda_2) (1-\sqrt{x_1}),
\\ &&
z_{31}=\frac{\lambda_2(1-\lambda_4)-\lambda_4}
{\lambda_2(1-\lambda_4)+\lambda_4},\, z_1=2\lambda_3-1,
\\
&& \nonumber z_{32}=2\lambda_5-1,
\end{eqnarray}
the differential phase space can be expressed in the following form:
\begin{widetext}
\begin{eqnarray}\label{PS4N}
\nonumber D\Phi(1\to 4)&=&\frac{m^{3d-8}}{128} \,
\frac{\Omega_{d-1}\Omega_{d-2}\Omega_{d-3}}{(2\pi)^{3d-4}}
\,d\lambda_1 d\lambda_2d\lambda_3d\lambda_4d\lambda_5
\\
&\times& \nonumber
(1-\sqrt{x_1})^{3d-7}\left[(1-\lambda_1)((1+\sqrt{x_1})^2-\lambda_1(1-\sqrt{x_1})^2)\right]^{(d-3)/2}
\\&\times&
\left[\lambda_1(1-\lambda_2)\right]^{d-3}
\left[\lambda_2(1-\lambda_3)\lambda_3(1-\lambda_4)\lambda_4\right]^{(d-4)/2}
\left[\lambda_5(1-\lambda_5)\right]^{(d-5)/2}.
\end{eqnarray}
\end{widetext}
where $\lambda_i=[0,1]$, $i=1...5$. The corresponding
scalar products in terms of the $\lambda_i$ will have the following
form
\begin{eqnarray}
\nonumber s_{234}&=&(p_2+p_3+p_4)^2=(1-\sqrt{x_1})^2 \lambda_1,
\\
\nonumber s_{34}&=&(p_3+p_4)^2=(1-\sqrt{x_1})^2 \lambda_1 \lambda_2,
\\
\nonumber s_{23}&=&(p_2+p_3)^2=(1-\sqrt{x_1})^2 \lambda_1
(1-\lambda_2) \lambda_4,
\end{eqnarray}
\begin{eqnarray}
\nonumber
s_{134}&=&(p_1+p_3+p_4)^2=(s_{134}^+-s_{134}^-)\lambda_3+s_{134}^-,
\\
s_{13}&=&(p_1+p_3)^2=(s_{13}^+-s_{13}^-)\lambda_5+s_{13}^-,
\end{eqnarray}
where
\begin{widetext}
\vspace{-0.2cm}
\begin{eqnarray}
\nonumber && \hspace{-0.4cm}
s_{134}^{\pm}=\frac{1}{2}\left[1+\lambda_2-\lambda_1(1-\lambda_2)(1-\sqrt{x_1})^2+x_1(1-\lambda_2)
\pm
(1-\sqrt{x_1})(1-\lambda_2)\sqrt{(1-\lambda_1)((1+\sqrt{x_1})^2-\lambda_1(1-\sqrt{x_1})^2)}\right],
\\
\nonumber && \hspace{-0.4cm}
s_{13}^{\pm}=x_1+\frac{1}{2}\,(1-\sqrt{x_1})\left[\left(\lambda_2\left(1-\lambda_4\right)+\lambda_4\right)
\left(1+\sqrt{x_1}-\lambda_1(1-\sqrt{x_1})\right)\right.
\\
\nonumber && \hspace{0.2cm}
+\left.\left(\lambda_2(1-\lambda_4)-\lambda_4\right)\left(1-2\lambda_3\right)\sqrt{\left(1-\lambda_1\right)
\left((1+\sqrt{x_1})^2-\lambda_1(1-\sqrt{x_1})^2\right)}\right]
\\
&& \hspace{0.2cm} \pm
2(1-\sqrt{x_1})\sqrt{\lambda_2(1-\lambda_3)\lambda_3(1-\lambda_4)\lambda_4(1-\lambda_1)
\left((1+\sqrt{x_1})^2-\lambda_1(1-\sqrt{x_1})^2\right)}\,.
\end{eqnarray}
\end{widetext}
As it is expected if we put $x_1=0$ in (\ref{PS4N}) we get the
formula (21) of \cite{Anastasiou:2003gr} derived for massless
particles in the final state.

In the Appendix B we consider the final state mass effects in some
interesting phenomenological cases, providing fully integrated
analytic formulae for the $1\to 4$ processes.

\newsection{Contribution to $B-\bar{B}$ mixing
at $\mathcal{O}(\alpha_s^2)$}

In this Section we illustrate how one can use formulae derived in
previous sections for calculation of  the absorptive part of a
$B-\bar{B}$ mixing at $\mathcal{O}(\alpha_s^2)$, which contributes
to the width differences $\Delta \Gamma$ and the CP-asymmetry.

An example of a diagram that contributes to the $B-\bar{B}$ mixing
at $O(\alpha_s^2)$ is shown in Fig.~1. This diagram contains both
3-particle and 4-particle cuts. The light quark masses we put equal
to zero.

We introduce a small mass $m_g$ for a gluon as an infrared
regulator. In this way, in the matching one doesn't need $\epsilon$
and $\epsilon^2$ parts of NLO and LO Wilson coefficients
correspondingly. Moreover, in the Effective Theory side the
renormalization is much easier, because the $(1/\epsilon^n)$ terms
appear only as UV singularities and so their cancelation can be
tracked easier. Further in order to eliminate IR finite terms coming
from the interference of IR $(1/\epsilon)$ with UV structures of
$O(\epsilon)$ a special class of operators $(\propto\epsilon)$ has
to be introduced in  case of dimensional regularization. Whereas
with introduction of a gluon mass those terms vanish automatically.
This is the major difference in using dimensional regularization or
gluon mass for IR singularities.

The reduction to the master integrals is done by means of a program
FIRE \cite{Smirnov:2008iw}.

Here we consider one of the most complicated master integrals (MI),
shown on the right side of the Fig.~1. This MI has one
three-particle cut and one four-particle cut. While in the case of
3-particle cut $i$ on the right side of the Fig.~1 denotes either
heavy $m_c$, $m_b$ or  light quarks, in the case of a 4-particle cut
$i$  stands for only  light quarks. The corresponding cuts can be
calculated analytically as an expansion over a gluon mass. The
coefficient of a corresponding MI contains inverse masses of $m_g$
up to the forth power.
\begin{figure}[t]\label{fig:PS_pic}
\vspace*{-4.8cm}
\begin{center}
\hspace*{-2.2cm}
\includegraphics[width=15cm,angle=0]{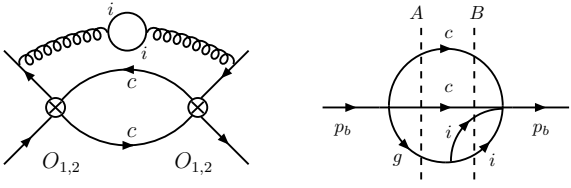}
\vspace*{-13.5cm} \caption{Left: an example of an infrared divergent
diagram contributing to the $B-\bar{B}$ process at NNLO.
 The virtual gluons are represented by curly lines. Right: one of the
 master integrals appearing after reduction of the left diagram with two
 cuts A and B ($c$- is a charm, $g$- is a gluon and $i$- is either charm or
  beauty or a light quark).}
\end{center}
\vspace*{-3.1mm}
\end{figure}

The 3-particle cut (cut A) can be factorized into a product of a
$q\bar{q}$ fermion loop and a 3-particle phase space, which is
presented in the Appendix, formula (\ref{PS3mmml}). In the case of
4-particle cut we get:
\begin{widetext}
\begin{eqnarray}
r_B &=&n_l\frac{m_b^2}{98304 \pi ^5} \left[16\pi^2(2-3x_c)x_c+192
x_c^2 \text{Li}_2(\sigma )-96 (2-x_c) x_c \text{Li}_2\left(\sigma
^2\right)\right.
   \\
   && \nonumber \hspace{1.8cm}
   +12 \sqrt{1-4 x_c} (2 x_c+1) (8 \log (1-\sigma )-4 \log (\sigma )-5)
      \\
&& \nonumber +48 \log (\sigma ) \left(x_c^2 \log (\sigma )-4 (2-x_c)
x_c \log (1-\sigma )+(1-x_c)^2+2 (2-x_c) x_c \log (1-4 x_c)\right)
\\
   && \nonumber +x_g\cdot \left(-8 \pi ^2 (5-6 x_c)
   +96 (1-2 x_c) \text{Li}_2(\sigma )+48 (3-2 x_c) \text{Li}_2\left(\sigma ^2\right)\right.
   \\
   && \nonumber \hspace{1cm} +24 \log (\sigma ) (4 (3-2 x_c) \log (1-\sigma )
   +(1-2 x_c) \log (\sigma )-2 (3-2 x_c) \log (1-4 x_c))
\\
&& \left. \nonumber  \hspace{1cm} +8 \sqrt{1-4 x_c} \left(7\pi
^2-6-48 \log ^2(1-\sigma )-3 \log ^2(\sigma )+48 \log (1-\sigma )
\log (\sigma )\right) \right)
    \\
   && \nonumber + x_g^2\cdot\frac{4}{(1-4 x_c)^{3/2}} \left(-28 \pi ^2 x_c^2-3 \left(6 x_c^2-10
   x_c+5\right)\right.
\\
&&  \nonumber \hspace{2cm} +12 \log (\sigma ) \left(1-2 x_c-2
x_c^2-\sqrt{1-4 x_c} (1-x_c)+x_c^2 \log (\sigma )\right)
\\
&& \left. \nonumber \hspace{2cm} -24 \log (1-\sigma ) \left(1-2
x_c-2 x_c^2-8 x_c^2 \log (1-\sigma )+8 x_c^2 \log (\sigma )\right)
\right)
   \\
   && \nonumber +\log (x_g) \cdot \left(-24 \sqrt{1-4 x_c} (2 x_c+1)-96 x_c (1-x_c) \log (\sigma
   )\right.
   \\
   && \nonumber \hspace{2.2cm} +x_g \left(96 (1-x_c)\log (\sigma )+96 \sqrt{1-4 x_c} \left(2 \log (1-\sigma )-\log (\sigma )\right)\right)
   \\
   && \left. \nonumber \hspace{2.2cm} +x_g^2 \frac{24}{(1-4 x_c)^{3/2}} \left(1-2 x_c-2 x_c^2-16 x_c^2 \log (1-\sigma )+8 x_c^2 \log (\sigma )\right) \right)
   \\
  && \nonumber \left. + \log ^2(x_g) \cdot
\left(-x_g 24 \sqrt{1-4 x_c}+x_g^2\frac{48 x_c^2}{(1-4
x_c)^{3/2}}\right) \right]+\mathcal{O}\left(x_g^3\right),
\end{eqnarray}
where $n_l$ is the number of light quarks ($n_l=3$),
$x_c=(m_c/m_b)^2$, $x_g=(m_g/m_b)^2$ and $\sigma=\frac{1-\sqrt{1-4
x_c}}{1+\sqrt{1-4 x_c}}$.\\
\end{widetext}

In the expression (4.2) the infrared singularities  appear as
$\log^n(x_g)$, n=1,2. The infrared singularities of the diagram
shown on the left side of Fig.~1, together with the $\log^n(x_g)$-s
of the other diagrams get canceled in the matching with the
corresponding Effective Theory diagrams with $Q$, $\tilde{Q}_S$
insertions \cite{Lenz:2006hd,Beneke:1996gn}. It must be mentioned
also, that the diagrams both in the full and Effective Theory have
to be renormalized. In addition, in the Effective Theory
corresponding Evanescent operators have to be taken into account as
well \cite{Herrlich:1994kh,Beneke:1998sy}.

\newsection{Conclusions \label{sec:concl}}

To conclude, we have presented new formulae for computing $1\to 3$
and $1\to 4$ processes, with massive particles in the final states.
We have also obtained a factorized formula for the 4 particle phase
space with 1 massive and 3 massless particles in the final state. On
the example of $B-\bar{B}$ mixing we  demonstrate  the capability of
our technique for  NNLO calculations, considering one of the most
complicated master integrals. Further we  study the final state mass
effects in some interesting phenomenological cases providing fully
integrated analytic formulae for 3  and 4 particle phase spaces.

\newappendix{ACKNOWLEDGMENTS}

This work was supported by the State Committee of Science of Armenia
program number 11-1c014 and VolkswagenStiftung program number 86426.
A. H. was partially supported by the EU contract MRTN-CT-2006-035482
(Flavianet) in early stage of this work.

\newappendix{APPENDIX A: FULLY INTEGRATED 3-PARTICLE PHASE-SPACE \label{app:interm}}
\def\theequation{A.\arabic{equation}}

Here, as an application of the formula (\ref{PS3}) we present fully
integrated 3-particle phase space. For the case of $m_1=m_2$,
$m_3=m_{light}$ ($m_{light}\ll m_{1,2}$), expanding (\ref{PS3}) over
$m_{light}/m$ and $\epsilon$ and performing integration,
 we get in the $\overline{MS}$  scheme
\begin{widetext}
\begin{eqnarray}\label{PS3mmml}
&& PS_3(m_1,m_1,m_{light})=
\left(\frac{m^2}{\mu^2}\right)^{-2\epsilon}
\frac{m^2}{128\pi^3}\left[ \frac{1}{2}\sqrt{1-4x_1}(1+2x_1)+2
x_1\left(1-x_1\right) \log (\sigma)\right.
   \\
   && \nonumber \hspace{1.4cm} +x_{\ell} \left(-2 \left(1-x_1\right) \log (\sigma)+\sqrt{1-4x_1}
   \left(\log (x_{\ell})-4 \log (1-\sigma)+2 \log (\sigma)\right)\right)
   \\
&& \nonumber \hspace{1.4cm} +x_{\ell}^2 \frac{1}{2(1-4x_1)^{3/2}}
\left(-1+2 x_1+2 x_1^2-4 x_1^2 \left(\log (x_{\ell})-4 \log
(1-\sigma)+2 \log (\sigma)\right)\right)
     \\
&& \nonumber +~ \epsilon\cdot \left[ -\frac{4 \pi ^2}{3} (1-x_1)
x_1+4 (1-x_1) x_1 \left(\text{Li}_2(\sigma )+\text{Li}_2\left(\sigma
^2\right)\right)\right.
\\
&& \nonumber \hspace{1cm}+(1+2 x_1) \sqrt{1-4 x_1}
\left(\frac{13}{4}-\frac{1}{2} \log (1-4 x_1)-2 \log (1-\sigma
)+\log (\sigma )\right)
\\
&& \nonumber \hspace{1cm} - (1-x_1) \log (\sigma ) \left(1-5 x_1+6
x_1 \log (1-4 x_1)-12 x_1 \log (1-\sigma )+x_1 \log (\sigma )\right)
   \\
   && \nonumber +x_{\ell} \left(\frac{4 \pi ^2}{3} (1-x_1) -4 (1-x_1)
   \left(\text{Li}_2(\sigma )+\text{Li}_2\left(\sigma ^2\right)\right)\right.
    \\
    && \hspace{1cm} \nonumber +(1-x_1) \log (\sigma )
    \left(-6+6 \log (1-4 x_1)-12 \log (1-\sigma )+\log (\sigma
    )\right)
    \\
    && \nonumber \hspace{1cm} +\sqrt{1-4 x_1} \left(
   4 \log (1-\sigma ) (2 \log (1-\sigma )-2 \log (\sigma )-3)+(6-\log (\sigma )) \log
   (\sigma )\right.
   \\
   && \left. \left. \nonumber \hspace{1cm} +\log (1-4 x_1) (4 \log (1-\sigma )-2 \log (\sigma )-\log (x_{\ell}))-\frac{1}{2} \log
   ^2(x_{\ell})+3 \log (x_{\ell})-2 \pi ^2+2\right)\right)
   \\
   && \nonumber +x_{\ell}^2\,
   \frac{1}{(1-4x_1)^{3/2}}\left(4 \pi ^2 x_1^2+\frac{1}{4} \left(22 x_1^2-14 x_1+7\right)
   \right.
   \\
   && \nonumber \hspace{1cm} +\log (x_{\ell}) \left(x_1^2 \log (x_{\ell})+\frac{1}{2} \left(-6 x_1^2+2 x_1-1\right)+2 x_1^2 \log (1-4 x_1)\right)
   \\
   && \nonumber \hspace{1cm} -2 \log (\sigma ) \left(-x_1^2 \log (\sigma )+2 x_1^2-2 x_1+1-\sqrt{1-4 x_1} (1-x_1)\right)
\\
&& \nonumber \hspace{1cm} + \frac{1}{2} \log (1-4 x_1) \left(-16
x_1^2 \log (1-\sigma )+8 x_1^2 \log (\sigma )-2 x_1^2-2 x_1+1\right)
   \\
   && \nonumber \left.\left.\left. \hspace{1cm} +\log (1-\sigma )
   \left(4 \left(2 x_1^2-2 x_1+1\right)-16 x_1^2 \log (1-\sigma )
   +16 x_1^2 \log (\sigma )\right)\right)\right] \right]
   +\mathcal{O}\left(\epsilon^2,x_{\ell}^3\right)\,.
\end{eqnarray}
For the case $m_1>m_3$ and $m_2=0$ we have
\begin{eqnarray}
&& PS_3(m_1,0,m_3)=\frac{m^2}{128\pi^3}\left[ \frac{1}{2}
(1+x_1+x_3) \sqrt{1-2 (x_1+x_3)+(x_1-x_3)^2}\right.
\\
&& \nonumber \hspace{1cm} -2 (x_1+x_3-2 x_1 x_3) \coth
^{-1}\left(\sqrt{\frac{1-\left(\sqrt{x_1}-\sqrt{x_3}\right)^2}
{1-\left(\sqrt{x_1}+\sqrt{x_3}\right)^2}}\right)
\\
&& \nonumber +(x_1-x_3) \left(2 \log
\left(\sqrt{1-\left(\sqrt{x_1}+\sqrt{x_3}\right)^2}
\left(\sqrt{x_1}-\sqrt{x_3}\right)+\sqrt{1-\left(\sqrt{x_1}-\sqrt{x_3}\right)^2}
   \left(\sqrt{x_1}+\sqrt{x_3}\right)\right)\right.
   \\
   && \left.\left. \nonumber \hspace{3cm} -\frac{1}{2} \log (16 x_1 x_3)\right)\right]
   +\mathcal{O}\left(\epsilon\right)\,.
\end{eqnarray}
For $m_1\gg m_3=m_{light}$, expanding over $m_{light}/m$, we get
also the terms proportional to $\epsilon^1$
\begin{eqnarray}
&& PS_3(m_1,0,m_{light})=
\left(\frac{m^2}{\mu^2}\right)^{-2\epsilon}
\frac{m^2}{128\pi^3}\left[ \frac{1}{2}
\left(1-x_1^2\right)+x_1\log(x_1) \right.
\\
&&\nonumber
+x_{\ell}\,((1-x_1)\log(x_{\ell})-2(1-x_1)\log(1-x_1)-x_1\log (x_1))
+x_{\ell}^2\left(\frac{1}{2}-\frac{1}{1-x_1}\right)
   \\
&& \nonumber +\epsilon\cdot\left[\frac{13}{4}
\left(1-x_1^2\right)-\frac{\pi^2x_1}{3}+\frac{1}{2}x_1(6-x_1-\log
(x_1))\log(x_1)\right.
\\
\nonumber && \hspace{1cm} -2\log(1-x_1)\left(1-x_1^2+x_1
\log(x_1)\right) -2x_1(\text{Li}_2(1-x_1)-\text{Li}_2(x_1))
\\
\nonumber && +x_{\ell} \left( (1-x_1) \left(6 \log ^2(1-x_1)-6 \log
(1-x_1)+3 \log (x_{\ell})+2\right)\right.
\\
&& \nonumber \hspace{1cm} -\frac{1}{2} (1-x_1) \left(4 \log (1-x_1)
\log (x_{\ell})+\log ^2(x_{\ell})\right) -\pi ^2 \left(1-\frac{4
x_1}{3}\right)
\\
&& \left. \nonumber \hspace{1cm} -x_1 \left(-2 \log
(1-x_1)-\frac{\log (x_1)}{2}+3\right) \log (x_1)+2
(\text{Li}_2(1-x_1)-x_1 \text{Li}_2(x_1))\right)
\\
&& \nonumber \left.\left. +x_{\ell}^2 \left(\frac{1+x_1}{1-x_1}
\left(3 \log (1-x_1)-\frac{\log
(x_{\ell})}{2}+\frac{7}{4}\right)-\frac{x_1 \log
(x_1)}{1-x_1}\right)\right]\right]+\mathcal{O}\left(\epsilon^2,x_{\ell}^3\right)\,,
\end{eqnarray}
where the following notations were introduced
\begin{eqnarray}\label{sigma}
x_{\ell}=(m_{light}/m)^2\,,\hspace{0.5cm}
x_{1,3}=(m_{1,3}/m)^2\,,\hspace{0.5cm} \sigma=\frac{1-\sqrt{1-4
x_1}}{1+\sqrt{1-4 x_1}}\,.
\end{eqnarray}
\end{widetext}

\newappendix{APPENDIX B: FULLY INTEGRATED 4-PARTICLE PHASE-SPACE}
\def\theequation{B.\arabic{equation}}

As an application of the formula (\ref{PS4}) we give fully
integrated 4-particle phase space. For the case when $m_1=m_2$,
$m_3=m_{light}$, $m_4=0$ ($m_{light}\ll m_{1,2}$), expanding
(\ref{PS4})  over $m_{light}/m$ and $\epsilon$ and performing
integration we get
\begin{widetext}
\begin{eqnarray}
&& \nonumber PS_4(m_1,m_1,m_{light},0)=\frac{m^4}{8192 \pi ^5}
\left[-\frac{4 \pi ^2 x_1^2}{3} +\frac{1}{3} \sqrt{1-4 x_1}
\left(1+20x_1+12 x_1^2\right)+4 x_1 \left(1+x_1-2 x_1^2\right) \log
(\sigma )\right.
\\
&& \hspace{1.3cm} +4 x_1^2 \log (\sigma ) (2 \log (1-4x_1)-4 \log
(1-\sigma )+\log (\sigma ))- 16 x_1^2 \text{Li}_2(-\sigma)
\\
&& \nonumber \hspace{0.5cm} + x_{\ell}\cdot \left(-16 x_1^2
\text{Li}_2(\sigma )+8x_1 (2-x_1) \text{Li}_2\left(\sigma
^2\right)+4 \left(x_1^2-1+\sqrt{1-4 x_1} (2x_1+1)\right) \log
(\sigma )\right.
\\
   && \nonumber\hspace{1.3cm} -4 x_1^2 \log ^2(\sigma )-\log (1-\sigma ) \left(8 \sqrt{1-4 x_1} (2 x_1+1)-16 (2-x_1) x_1
   \log (\sigma )\right)
\\
   && \nonumber \hspace{1.3cm}-8 (2-x_1) x_1 \log (\sigma ) \log (1-4
   x_1)+\log (x_{\ell}) \left(8 (1-x_1) x_1 \log (\sigma )+2 \sqrt{1-4
   x_1} (2 x_1+1)\right)
\\
   && \nonumber \left. \hspace{1.3cm}-\frac{4}{3} \pi ^2 (2-3 x_1) x_1+3 \sqrt{1-4 x_1}
   (2 x_1+1)\right)
\\
&& \nonumber \hspace{0.5cm} \left. + x_{\ell}^2\cdot \left(-4
(1-x_1) \log (\sigma )-\sqrt{1-4 x_1}\, (3+8 \log (1-\sigma )-4 \log
(\sigma)-2 \log
   (x_{\ell}))\right)\right]+\mathcal{O}\left(\epsilon,x_{\ell}^3\right)\,.
\end{eqnarray}
For $m_1>m_3$, $m_2=m_4=0$ we have
\begin{eqnarray} \label{ff2}
&& \nonumber PS_4(m_1,0,m_3,0)=\frac{m^4}{8192 \pi ^5} \left[4
\left(x_1^2-x_3^2\right) \tanh
   ^{-1}\left(\sqrt{\frac{1-\left(\sqrt{x_1}+\sqrt{x_3}\right)^2}
   {1-\left(\sqrt{x_1}-\sqrt{x_3}\right)^2}}
   \frac{\sqrt{x_1}-\sqrt{x_3}}{\sqrt{x_1}+\sqrt{x_3}}\right)\right.
\\
&& \hspace{1.0cm} +\frac{1}{3} \sqrt{1-2 (x_1+x_3)+(x_1-x_3)^2}
\left(1+x_1^2+x_3^2+10 (x_1+x_3+x_1 x_3)\right)
   \\
   && \nonumber \hspace{1.0cm}+2 (x_1 - x_3) \log (4 x_1)+\left(x_1^2 (1-2 x_3)+2 x_3 (1-x_1 x_3)+x_3^2+4 x_1 x_3 \log (x_3)\right)
      \\
   && \nonumber \hspace{1.2cm} \times \left(\log (4 x_1 x_3)-2 \log \left(1-x_1-x_3+\sqrt{1-2
(x_1+x_3)+(x_1-x_3)^2}\right)\right)
   \\
   && \nonumber \hspace{1.0cm}-4 (x_1-x_3) \log \left(1+x_1-x_3+\sqrt{1-2
(x_1+x_3)+(x_1-x_3)^2}\right)
   \\
   && \nonumber \hspace{1.0cm}-8 x_1 x_3 \text{Li}_2\left(\frac{2 x_1}{x_1+x_3-1-\sqrt{1-2
(x_1+x_3)+(x_1-x_3)^2}}\right)
   \\
   && \nonumber \left. \hspace{1.0cm}+8 x_1 x_3 \text{Li}_2\left(\frac{x_1+x_3-1-\sqrt{1-2
(x_1+x_3)+(x_1-x_3)^2}}{2
   x_3}\right)\right]+\mathcal{O}\left(\epsilon\right)\,.
\end{eqnarray}
For $m_1\gg m_3=m_{light}$, expanding the expression (\ref{ff2})
over $m_3$, we get
\begin{eqnarray}
&& PS_4(m_1,0,m_{light},0)=\frac{m^4}{8192 \pi ^5} \left[\frac{1}{3}
(1-x_1) \left(x_1^2+10x_1+1\right)+2 x_1 (x_1+1) \log (x_1)\right.
\\
&& \nonumber \hspace{0.5cm} +x_{\ell} \left(\left(1-x_1^2\right)
(3-4 \log (1-x_1)+2 \log (x_{\ell})) +2 x_1 \log (x_1) (-x_1-4 \log
(1-x_1)+2 \log (x_{\ell})) \right.
\\
&& \left. \nonumber  \hspace{1.3cm}-8 x_1 \text{Li}_2(1-x_1)\right)
\\
&& \nonumber  \hspace{0.5cm} \left.+x_{\ell}^2 (-(1-x_1) (3+4 \log
(1-x_1)-2 \log (x_{\ell}))-2 x_1 \log
(x_1))\right]+\mathcal{O}\left(\epsilon,x_{\ell}^3\right)\,.
\end{eqnarray}
\end{widetext}

\end{document}